\documentclass[showpacs,twocolumn]{revtex4}

\usepackage{ragged2e}
\usepackage{amsmath}
\usepackage{graphicx}
\usepackage[usenames,dvipsnames]{color}

\definecolor{darkblue}{RGB}{0,0,196}

\begin{document}

\title{\Large \bf Initial- and final-state temperatures of emission source
from differential cross-section in squared momentum transfer in
high energy collisions\vspace{0.5cm}}

\author{Qi Wang$^{1,}$\footnote{qiwang-sxu@qq.com; 18303476022@163.com}}

\author{Fu-Hu Liu$^{1,}$\footnote{Correspondence: fuhuliu@163.com;
fuhuliu@sxu.edu.cn}}

\author{Khusniddin K. Olimov$^{2,}$\footnote{Correspondence: khkolimov@gmail.com; kh.olimov@uzsci.net}}

\affiliation{$^1$Institute of Theoretical Physics \& Collaborative
Innovation Center of Extreme Optics \& State Key Laboratory of \\
Quantum Optics and Quantum Optics Devices, Shanxi University,
Taiyuan 030006, China}

\affiliation{$^2$Laboratory of High Energy Physics,
Physical-Technical Institute of SPA ``Physics-Sun" of Uzbek
Academy of Sciences, \\ Chingiz Aytmatov str. $2^b$, Tashkent
100084, Uzbekistan}

\begin{abstract}

\vspace{0.5cm}

\noindent The differential cross-section in squared momentum
transfer of $\rho$, $\rho^0$, $\omega$, $\phi$, $f_{0}(980)$,
$f_{1}(1285)$, $f_{0}(1370)$, $f_{1}(1420)$, $f_{0}(1500)$, and
$J/\psi$ produced in high energy virtual photon-proton
($\gamma$$^{*} p$), photon-proton ($\gamma p$), and proton-proton
($pp$) collisions measured by the H1, ZEUS, and WA102
Collaborations are analyzed by the Monte Carlo calculations. In
the calculations, the Erlang distribution, Tsallis distribution,
and Hagedorn function are separately used to describe the
transverse momentum spectra of the emitted particles. Our results
show that the initial- and final-state temperatures increase from
lower squared photon virtuality to a higher one, and decrease with
the increase of center-of-mass energy.

\end{abstract}
\pacs{12.40.Ee, 14.40.-n, 24.10.Pa, 25.75.Ag \vspace{0.5cm}}

\maketitle

\section{Introduction}

In high-energy collisions, it is interesting for us to describe
the excitation and equilibrium degrees of an interacting system
because of the two degrees related to the reaction mechanism and
evolution process of the collision
system~\cite{1a,1b,1c,1d,1e,1f,1g,1h,1i,1j}. In the progress of
describing the excitation degree and structure character of the
system, temperature is an important quantity in physics in view of
intuitiveness and representation. In high-energy collisions,
different types of temperatures are used~\cite{1,2,3,4,5,6,6a,6b},
which usually refer to the initial-state temperature $T_i$,
quark-hadron transition temperature $T_{tr}$, chemical freeze-out
temperature $T_{ch}$, kinetic freeze-out or final-state
temperature (``confinement" temperature) $T_{kin}$ or $T_0$, and
effective temperature $T_{eff}$ or $T$, etc. In this work, we
emphatically discuss the initial- and final-state temperatures,
though other types of temperatures are also important.

The initial temperature $T_i$ is the temperature of emission
source or interacting system when a projectile particle or nucleus
and a target particle or nucleus undergo initial stage of a
collision. It represents the excitation degree of emission source
or that of an interacting system in the initial state of
collisions, and it is usually meant as describing the interacting
system after thermalization. The initial temperature $T_i$ can be
extracted by fitting the transverse momentum $p_T$ spectra of
particles by using some distributions such as the Erlang
distribution~\cite{7,8,9}, Tsallis distribution~\cite{11,12},
Hagedorn function~\cite{13}, L{\'e}vy--Tsallis function~\cite{14}.
Here, both the names of distribution and function are used
according to the various accepted terminology in the literature,
though they represent the similar probability density function in
fitting the particle spectra. Meanwhile, the average transverse
momentum $\langle p_T\rangle$ can be obtained from the same
function.

The final-state temperature $T_0$ is usually known as the kinetic
freeze-out temperature, which refers to the temperature of
emission source when the inelastic collisions ceased and there are
only elastic collisions among particles. In the last stage of
collisions, the momentum distribution of particles is fixed and
the transverse momentum spectra can be measured in experiments.
The excitation degree of the system in the last stage can be
described by the final-state temperature $T_0$ in which the
influence of flow effect is excluded. The temperature or related
main parameters used in the Erlang distribution~\cite{7,8,9},
Tsallis distribution~\cite{11,12}, Hagedorn function~\cite{13},
and L{\'e}vy--Tsallis function~\cite{14} are not $T_0$, but the
effective temperature $T$ in which the influence of flow effect is
not excluded.

The Mandelstam variables~\cite{10} consist of the four-momentum of
particles in two-body reaction. Both the squared momentum transfer
and transverse momentum can represent the kinetic character of
particles. Let us use the squared momentum transfer to replace
transverse momentum in fitting the particle spectra. Then, we can
fit the squared momentum transfer spectra with the related
distributions to obtain the initial temperature $T_i$, average
transverse momentum $\langle p_T\rangle$, and other quantities. Of
course, in fitting the squared momentum transfer spectra, the
above mentioned distributions cannot be used directly. In fact, we
have to use the Monte Carlo method to obtain the concrete value of
a transverse momentum for a given particle from the mentioned
distributions. Then, the concrete value of squared momentum
transfer can be obtained from the definition.

Except the temperature parameter, other parameters also describe
partly the characters of the interacting system. For instance, the
entropy index $q$ which describes the degree of equilibrium can be
extracted from the Tsallis distribution~\cite{11,12} considering
the particle mass. Meanwhile, $q$ can be extracted from the
Hagedorn function~\cite{13} which is the same as the
L{\'e}vy--Tsallis function~\cite{14} for a particle neglecting its
mass. If there is relation between the Tsallis distribution and
Hagedorn function, we may say that the former one covers the
latter one in which the mass is neglected. Because the
universality, similarity or common characteristics exist in high
energy collisions~\cite{14a,14b,14c,14d,14e,14f,14g,14h,14i,14j},
some distributions used in large collision system can be also used
in small collision system.

In this paper, the differential cross-section in squared momentum
transfer of $\rho$, $\rho^0$, $\phi$, and $J/\psi$ produced in
virtual photon-proton ($\gamma$$^{*} p$) collisions, $\omega$ and
$J/\psi$ produced in photon-proton ($\gamma p$) collisions, as
well as $f_{0}(980)$, $f_{1}(1285)$, $f_{0}(1370)$, $f_{1}(1420)$,
and $f_{0}(1500)$ produced in proton-proton ($pp$) collisions
measured by the H1~\cite{15,22}, ZEUS~\cite{16,17,18,21}, and
WA102 Collaborations~\cite{19,20} are fitted with the results from
the Monte Carlo calculations. Firstly, the transverse momenta
satisfied with the Erlang distribution, Tsallis distribution, and
Hagedorn function are generated. Secondly, these special
transverse momenta are transformed to the squared momentum
transfers. Thirdly and lastly, the distribution of squared
momentum transfers is obtained and fitted to the experimental data
by the least squares method.

\section{Formalism and method}

i) {\it The Erlang distribution}

The Erlang distribution is the convolution of multiple exponential
distributions. In the framework of multi-source thermal
model~\cite{7,8,9}, we may think that more than one parton (or
parton-like) contribute to the transverse momentum of considered
particle. The $j$-th parton (or parton-like) is assumed to
contribute to the transverse momentum to be $p_{tj}$ which obeys
an exponential distribution with the average $\langle p_t\rangle$
which is $j$-ordinal number independent. We have the probability
density function obeyed by $p_{tj}$ to be
\begin{align}
f(p_{tj})=\frac{1}{\langle p_t \rangle}
\exp\bigg(-\frac{p_{tj}}{\langle p_t \rangle}\bigg).
\end{align}
The average $\langle p_t\rangle$ reflects the excitation degree of
contributor parton and can be regarded as the effective
temperature $T$.

The contribution of all $n_s$ partons to $p_T$ is the sum of
various $p_{tj}$. The distribution of $p_T$ is then the
convolution of $n_s$ exponential functions~\cite{7,8,9}. We have
the $p_T$ distribution (the probability density function of $p_T$)
of final-state particles to be the Erlang distribution
\begin{align}
f_1(p_T)=\frac{1}{N}\frac{dN}{dp_T}=\frac{p_T^{n_s-1}}{(n_s-1)!{\langle
p_t \rangle}^{n_s}} \exp\bigg(-\frac{p_T}{{\langle p_t
\rangle}}\bigg),
\end{align}
where $N$ denotes the number of all considered particles and $p_T$
has an average of $\langle
p_T\rangle=\int_0^{\infty}p_Tf_1(p_T)dp_T=n_s\langle p_t\rangle$.
Eq. (2) is naturally normalized to be 1. In Eq. (2), there are two
free parameters, $n_s$ and $\langle p_t\rangle$.
\\

ii) {\it The Tsallis distribution}

The Tsallis distribution~\cite{11,12} has more than one form,
which are widely used in the field of high energy collisions.
Conveniently, we use the following form
\begin{align}
f_2(p_T)=\frac{1}{N}\frac{dN}{dp_T}=C p_T
\bigg(1+\frac{m_T-m_0}{nT}\bigg)^{-n},
\end{align}
where $C$ is the normalization constant, $m_T =
\sqrt{p_T^2+m_0^2}$ is the transverse mass, $m_0$ is the rest
mass, $n=1/(q-1)$, and $q$ is the entropy index~\cite{11,12}. Eq.
(3) is valid only at mid-rapidity ($y\approx 0$) which results in
$\cosh y \approx 1$ and the particle energy $E=m_T\cosh y \approx
m_T$.

In Eq. (3), a large $n$ corresponds to a $q$ that is close to 1,
and the source or system approaches to equilibrium. The larger the
parameter $n$ is, the closer to 1 the entropy index $q$ is, with
the source or system being at a higher degree of equilibrium.
There is not exactly the minimum $n$ or maximum
$q$~\cite{11,12,13,14} which is a limit for approximate
equilibrium. Empirically, in the case of $n\ge4$ or $q\le1.25$
which is 25\% more than 1 (even $n\ge5$ or $q\le1.2$ which is 20\%
more than 1), the source or system can be regarded as being in a
state of approximate (local) equilibrium. Usually, in high energy
collisions, the source or system is approximately in equilibrium
due to $n$ being large enough.
\\

iii) {\it The Hagedorn function}

The Hagedorn function~\cite{13} is an inverse power-law which has
the probability density function of $p_T$ to be
\begin{align}
f_3(p_T)=\frac{1}{N}\frac{dN}{dp_T}=Ap_T\bigg(1+\frac{p_T}{p_0}
\bigg)^{-n_0},
\end{align}
where $A$ is the normalization constant, $n_0$ is a free parameter
which is similar to $n$ in the Tsallis distribution~\cite{11,12},
and $p_0$ is a free parameter which is similar to the product of
$nT$ in the Tsallis distribution. Note here that it appears as
$p_0=nT$ is a perfect liquid like relation; however, $p_0$ is
transverse momentum, and $n$ is a dimensionless number. This is
not meant in a perfect liquid sense, but the letters are just
randomly coinciding.

It should be noted that the Hagedorn function is a special case of
the Tsallis distribution in which $m_0$ can be neglected.
Generally, at high $p_T$, we may neglect $m_0$, observing the two
distributions being very similar to each other. At low $p_T$, the
two distributions have obvious differences due to non-ignorable
$m_0$. To build a connection with the entropy index $q$, we have
$n_0\approx 1/(q-1)$. To build a connection with the effective
temperature $T$, we have $p_0\approx n_0T \approx T/(q-1)$.
\\

iv) {\it The squared momentum transfer}

In the center-of-mass reference frame, in two-body reaction
$2+1\rightarrow 4+3$ (e.g. $\gamma^*p\rightarrow \rho p$) or in
two-body-like reaction, it is supposed that particle 1 is incident
along the $z$ direction and particle 2 is incident along the
opposite direction. In addition, particle 3 is emitted with angle
$\theta$ relative to the $z$ direction and particle 4 is emitted
along the opposite direction. According to ref.~\cite{10}, three
Mandelstam-variables are defined as
\begin{align}
s=-({P_1}+{P_2})^{2}=-({P_3}+{P_4})^{2},
\end{align}
\begin{align}
t=-({P_1}-{P_3})^{2}=-(-{P_2}+{P_4})^{2},
\end{align}
\begin{align}
u=-({P_1}-{P_4})^{2}=-(-{P_2}+{P_3})^{2},
\end{align}
where $P_{1}$, $P_{2}$, $P_{3}$, and $P_{4}$ are four-momenta of
particles 1, 2, 3, and 4, respectively.

In the Mandelstam-variables, slightly varying the form, $\sqrt{s}$
is the center-of-mass energy, $-t$ is the squared momentum
transfer between particles 1 and 3, and $-u$ is the squared
momentum transfer between particles 1 and 4. Conveniently, let
$|t|$ be the squared momentum transfer between particles 1 and 3.
We have
\begin{align}
|t|=& |({E_1}-{E_3})^{2}-({\vec{p}_{1}}-{\vec{p}_{3}})^{2}| \nonumber\\
=& \bigg|m_1^2+m_3^2-2{E_1}\sqrt{\bigg({\frac{p_{3T}}{\sin\theta}}\bigg)^{2}+m_3^2} \nonumber\\
& +2\sqrt{E_1^2-m_1^2}\frac{p_{3T}}{\tan\theta}\bigg|.
\end{align}
Here $E_1$ and $E_3$, $\vec{p}_1$ and $\vec{p}_3$, as well as
$m_1$ and $m_3$ are energy, momentum, and rest mass of particles 1
and 3, respectively. In particular, $p_{3T}$ is the transverse
momentum of particle 3, which is referred to be perpendicular to
the $z$ direction.

As the energy of incoming photon in the center-of-mass reference
frame of the reaction, $E_1$ in Eq. (8) should be a fixed value.
However, $E_1$ has a slight shift from the peak value due to
different experiments and selections. To obtain a good fit, we
treat $E_1$ as a parameter which is the same or has small
difference in the same/similar reactions. $p_{3T}$ obeys one of
Eqs. (2)--(4) and $\theta$ obeys an isotropic assumption in the
center-of-mass reference frame, which will be discussed later in
this section. To obtain $|t|$, we may perform the Monte Carlo
calculations. Note that we may calculate $|t|$ from two particles,
i.e., particles 1 and 3, but not from one particle. Instead, for
one calculation, $|t|$ means the squared momentum transfer in an
event. For many calculations, $|t|$ distribution can be obtained
from the statistics. For convenience in the description, the
transverse momentum and rest mass of particle 3 are also denoted
by $p_T$ and $m_0$ respectively.

Based on the experiments cited from
literature~\cite{15,22,16,17,18,21,19,20}, we have used two main
selection factors for the data. 1) The squared photon virtuality
$Q^2=-P_{\gamma}^2$, where $P_{\gamma}$ denotes the four-momentum
of photon. 2) The center-of-mass energy $\sqrt{s}$ or $W$, i.e.,
$W=\sqrt{s}=\sqrt{-({P_1}+{P_2})^{2}}$. Let $x$ denote the Bjorken
scaling variable, one has $W^2\simeq Q^2/x$.
\\

v) {\it The initial- and final-state temperatures}

According to refs.~\cite{23,24,25}, in a color string percolation
approach, the initial temperature $T_i$ can be estimated as
\begin{align}
T_i =\sqrt{\frac{\langle p_T^2 \rangle}{2}},
\end{align}
where $\sqrt{\langle p_T^2 \rangle}$ is the root-mean-square of
$p_T$ and $\langle p_T^2 \rangle=\int_0^{\max} p^2_T
f_{1,2,3}(p_T) dp_T$. In the expression of initial temperature, we
have used a single string in the cluster for a given particle
production~\cite{51}, though more than two partons or partons-like
take possibly part in the formation of the string. That is, we
have used the color suppression factor $F(\xi)$ to be 1 in the
color string percolation model~\cite{51}. Other strings, even if
they exist, do not affect noticeably the production of a given
particle. If other strings are considered, i.e., if we take the
minimum $F(\xi)$ to be 0.6~\cite{51}, a higher $T_i$ can be
obtained by multiplying a revised factor, $\sqrt{1/F(\xi)}=1.291$,
in Eq. (9).

The extraction of final-state temperature $T_0$ is more complex
than that of the initial temperature $T_i$. Generally, one may
introduce the transverse flow velocity $\beta_T$ in the considered
function and obtain $T_0$ and $\beta_T$
simultaneously~\cite{52,53,54,55,56,57,58,59,60}, in which the
effective temperature $T$ no longer appears. Alternatively, the
intercept in $T$ versus $m_0$ is assumed to be
$T_0$~\cite{53,61,62,63,64,65,66}, and the slope in $\langle
p_T\rangle$ versus $\overline{m}$ is assumed to be
$\beta_T$~\cite{65,66,67,68,69}, where $\overline{m}$ denotes the
average energy. However, the alternative method using intercept
and slope is not suitable for us due to the fact that the spectra
of more than two types of particles (e.g., pions, kaons, and
protons) are needed in the extraction which is not our case.

In $\gamma$$^{*} p$, $\gamma p$, and $pp$ collisions discussed in
the present work, the flow effect is not considered by us due to
the collective effect being small in the two-body process. This
means that $T_0\approx T$ in the considered processes. Here, $T$
appears as that in Eq. (3). Meanwhile, $T$ can be also
approximated by $\langle p_t\rangle$ in Eq. (1) and $p_0/n_0$ in
Eq. (4). Generally, we may regard different distributions or
functions as different ``thermometers". Just like the Celsius
thermometer and the Fahrenheit thermometer, different thermometers
measure different temperatures, though they can be transformed
from one to another according to conversion rules. Although we may
approximately regard $T$ in Eq. (3) as $T_0$, a smaller $T_0$ can
be obtained if the flow effect is considered.

As mentioned above, $T=\langle p_t\rangle=\langle p_T\rangle/n_s$
in Eq. (1) and Erlang distribution, and $T_i =\sqrt{\langle p_T^2
\rangle/2}$. We have $T^2n_s^2=2T_i^2-\sigma^2_{p_T}$, so this
would mean that $T$ is basically encoded in $\sigma^2_{p_T}$, the
squared variance of $p_T$ in the distribution. This also means
that $T_i$ and $T$ are related through $p_T$. It is
understandable, because they reflect the violent degrees of
collisions at different stages. Generally, $T_i>T$, this is
natural.

Note that although we may use the final-state temperature, it is
not a freeze-out temperature for the small system discussed in
this paper. In particular, for $\gamma^*p$ and $\gamma p$
reactions, these are just a process describable in terms of
perturbative quantum chromodynamics (pQCD) and
factorization~\cite{69a}, but not a process in which
de-confinement or freeze-out is involved. The meaning of
final-state temperature for the large system such as heavy ion
collisions or small system such as $pp$ collisions with high
multiplicity is somehow different from here. At least, for the
large system, we may consider the de-confinement or freeze-out
involved picture. Meanwhile, the flow effect in the large system
cannot be neglected.
\\

vi) {\it The process of Monte Carlo calculations}

In analytical calculation, the functions Eqs. (2)--(4) on $p_T$
distribution are difficult to be used in Eq. (8) to obtain $|t|$
distribution. Instead, we may perform the Monte Carlo
calculations. Let $R_{1,2}$ and $r_{1,2,3,...,n_s}$ be random
numbers distributed evenly in [0,1]. To use Eq. (8), we have to
know changeable $p_{3T}$ (i.e. $p_T$) and $\theta$. Other
quantities such as $E_1$, $m_1$, and $m_3$ in the equation are
fixed, though $E_1$ is treated by us as a parameter with slight
variety.

To obtain a concrete value of $p_T$, we need one of Eqs. (2)--(4).
Solving the equation
\begin{align}
\int_0^{p_T}f_i(p'_T)dp'_T<R_1<\int_0^{p_T+\delta
p_T}f_i(p'_T)dp'_T,
\end{align}
where $i=1$, 2, and 3, respectively, and $\delta p_T$ is a small
shift relative to $p_T$, we may obtain concrete $p_T$. It seems
that Eq. (10) directly means that the integral of $f_1(p_T)$,
$f_2(p_T)$, and $f_3(p_T)$ is the same for the $[0,p_T]$ interval,
which essentially means that the three functions are equal (except
for a null measure set). In fact, the three functions are
different in forms because of Eqs. (2)--(4), and we need to
distinguish them.

In particular, for $f_1(p_T)$, we have a simpler expression. Let
us solve the equation
\begin{align}
\int_0^{p_{tj}}f(p'_{tj})dp'_{tj}=r_j \quad (j=1,2,3,...,n_s).
\end{align}
We have
\begin{align}
p_{tj}=-\langle p_t\rangle \ln r_j \quad (j=1,2,3,...,n_s)
\end{align}
due to Eq. (1) being used, where $r_j$ in Eq. (12) replaced
$1-r_j$ because both of them are random numbers in [0,1]. The
simpler expression is
\begin{align}
p_T=-\langle p_t\rangle \ln\bigg(\prod_{j=1}^{n_s} r_j\bigg)
\end{align}
due to $p_T$ being the sum of $n_s$ random $p_{tj}$.

To obtain a concrete value of $\theta$, we need the function
\begin{align}
f_{\theta}(\theta)=\frac{1}{2}\sin\theta
\end{align}
which is obeyed by $\theta$ under the assumption of isotropic
emission in the center-of-mass reference frame. Solving the
equation
\begin{align}
\int_0^{\theta}f_{\theta}(\theta')d\theta'=R_2,
\end{align}
we have
\begin{align}
\theta=2\arcsin(\sqrt{R_2})
\end{align}
which is needed by us.

According to the concrete values of $p_T$ and $\theta$, and using
other quantities, the value of $|t|$ can be obtained from Eq. (8).
After repeating the calculations many times, the distribution of
$|t|$ is obtained statistically. Based on the method of least
squares, the related parameters are obtained naturally. Meanwhile,
$T_i$ can be obtained from Eq. (9). $\langle p_T\rangle$ and
$\langle p_T^2\rangle$ can be obtained from one of Eqs. (2)--(4)
or from the statistics. The errors of parameters are obtained by
the general method of statistical simulation.

\section{Results and discussion}

\subsection{Comparison with data}

Figure 1 shows the differential cross-section in squared momentum
transfer, $d\sigma/d|t|$, of (a) $\gamma^*p \rightarrow \rho p$,
(b) $\gamma^*p \rightarrow \rho Y$, and (c) $\gamma^*p \rightarrow
\rho^0p$ produced in electron-proton ($ep$) collisions at
photon-proton center-of-mass energy (a)(b) $W=75$ GeV and (c)
$W=90$ GeV, where $\sigma$ denotes the cross-section and $Y$ in
panel (b) denotes an ``elastic" scattering proton or a
diffractively excited ``proton dissociation"~\cite{15}. The
experimental data points from (a)(b) non-exclusive and (c)
exclusive productions are measured by the H1~\cite{15} and ZEUS
Collaborations~\cite{16}, respectively, with different average
squared photon virtuality (a) $Q^2=3.3$, 6.6, 11.5, 17.4, and 33.0
GeV$^2$, (b) $Q^2=3.3$, 6.6, and 15.8 GeV$^2$, as well as (c)
$Q^2=2.7$, 5.0, 7.8, 11.9, 19.7, and 41.0 GeV$^2$. The data points
are fitted by the Monte Carlo calculations with the Erlang
distribution Eq. (2) (the solid curves), the Tsallis distribution
Eq. (3) (the dashed curves), and the Hagedorn function Eq. (4)
(the dotted curves) for $p_{3T}$ in Eq. (8). Some data are scaled
by different quantities marked in the panels for clear visibility.
In the calculations, the method of least squares is used to obtain
the parameter values. The values of $E_1$, $\langle p_t \rangle$,
$n_s$, $T_i$, $T$, $n$, $p_0$, and $n_0$ are listed in Tables 1,
2, or 3 with $\chi^2$ and number of degree of freedom (ndof). One
can see that in most cases the calculations based on Eq. (8) with
Eqs. (2)--(4) for $p_{3T}$ can fit approximately the experimental
data measured by the H1 and ZEUS Collaborations.

\begin{figure*}[!htb]
\begin{center}
\includegraphics[width=13cm]{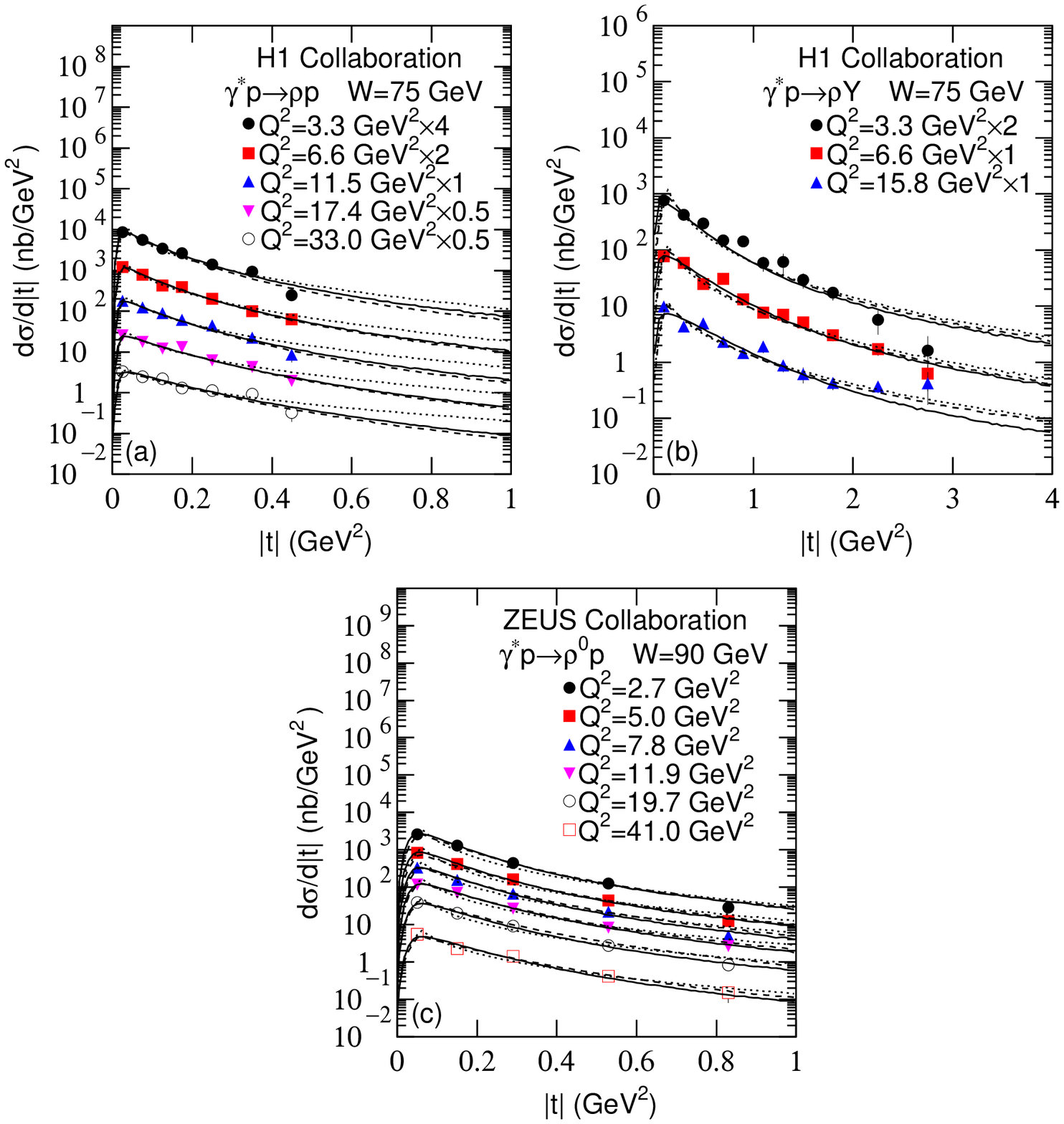}
\end{center}
\justifying\noindent {\small Fig. 1. The differential
cross-section in squared momentum transfer of (a) $\gamma^*p
\rightarrow \rho p$, (b) $\gamma^*p \rightarrow \rho Y$, and (c)
$\gamma^*p \rightarrow \rho^0p$ produced in $ep$ collisions at
(a)(b) $W=75$ GeV and (c) $W=90$ GeV. The experimental data points
from (a)(b) non-exclusive and (c) exclusive productions are
measured by the H1~\cite{15} and ZEUS Collaborations~\cite{16},
respectively, with different $Q^2$ marked in the panels. The data
points are fitted by the Monte Carlo calculations with the Erlang
distribution Eq. (2) (the solid curves), the Tsallis distribution
Eq. (3) (the dashed curves), and the Hagedorn function Eq. (4)
(the dotted curves) for $p_{3T}$ in Eq. (8).}
\end{figure*}

\begin{figure*}[!htb]
\begin{center}
\includegraphics[width=13cm]{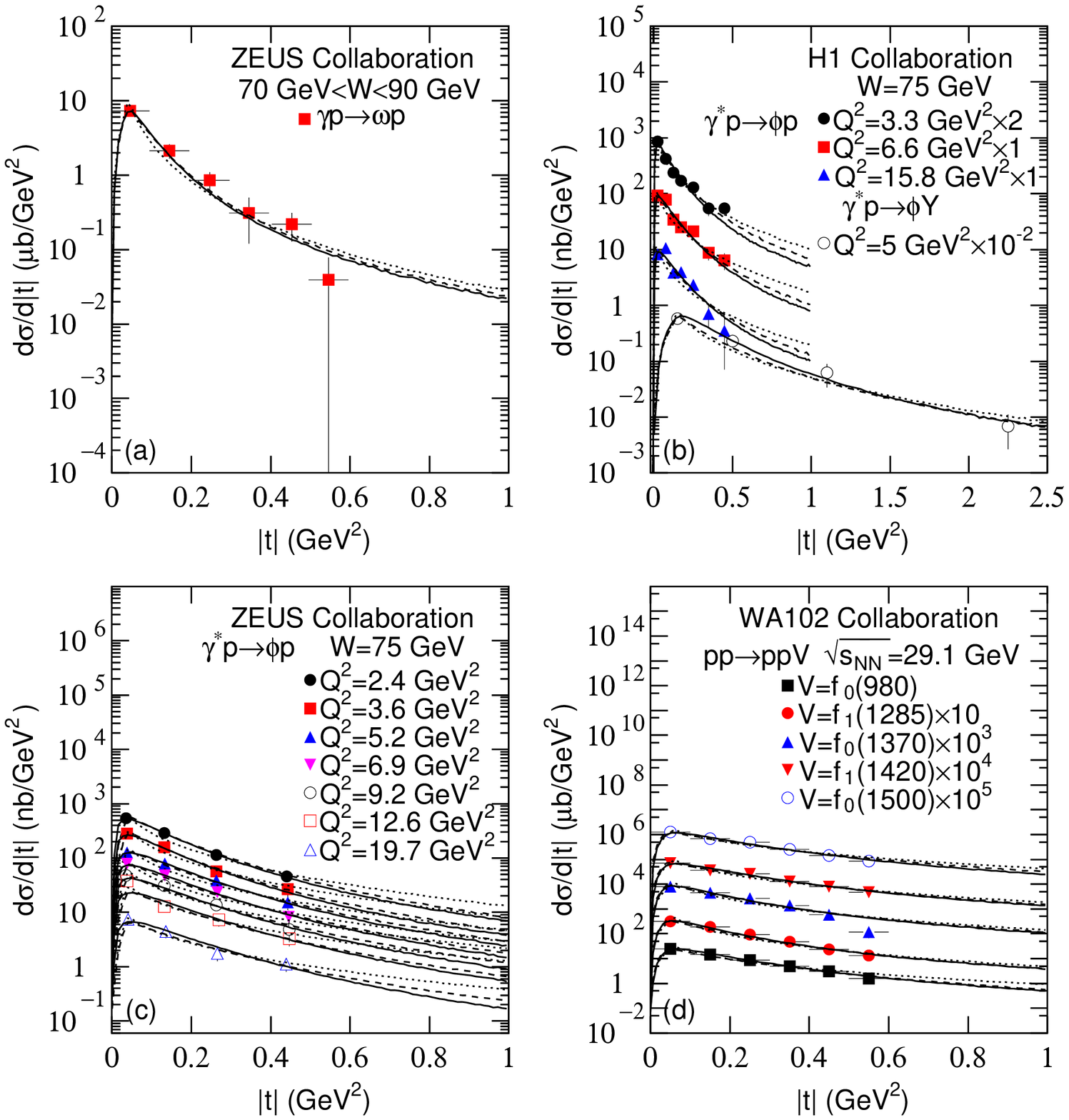}
\end{center}
\justifying\noindent {\small Fig. 2. The differential
cross-section in squared momentum transfer of (a) $\gamma p
\rightarrow \omega p$, (b) $\gamma^*p \rightarrow \phi p$ and
$\gamma^*p \rightarrow \phi Y$, (c) $\gamma^*p \rightarrow \phi
p$, and (d) $pp \rightarrow ppV$ ($V=f_0(980)$, $f_1(1285)$,
$f_0(1370)$, $f_1(1420)$, and $f_0(1500)$) produced in (a)--(c)
$ep$ and (d) $pp$ collisions in (a) 70 ${\rm GeV}<W<90$ GeV, at
(b)(c) $W=75$ GeV, and at (d) $\sqrt{s_{NN}}=29.1$ GeV. The
experimental data points from (a)(c) exclusive, (b) non-exclusive,
and (d) exclusive productions are measured by the
ZEUS~\cite{17,18}, H1~\cite{15}, and WA102
Collaborations~\cite{19,20}, respectively, with different $Q^2$
for only panels (b) and (c). Similar to Figure 1, the data points
are fitted by the Monte Carlo calculations based on Eq. (8).}
\end{figure*}

\begin{figure*}[!htb]
\begin{center}
\includegraphics[width=13cm]{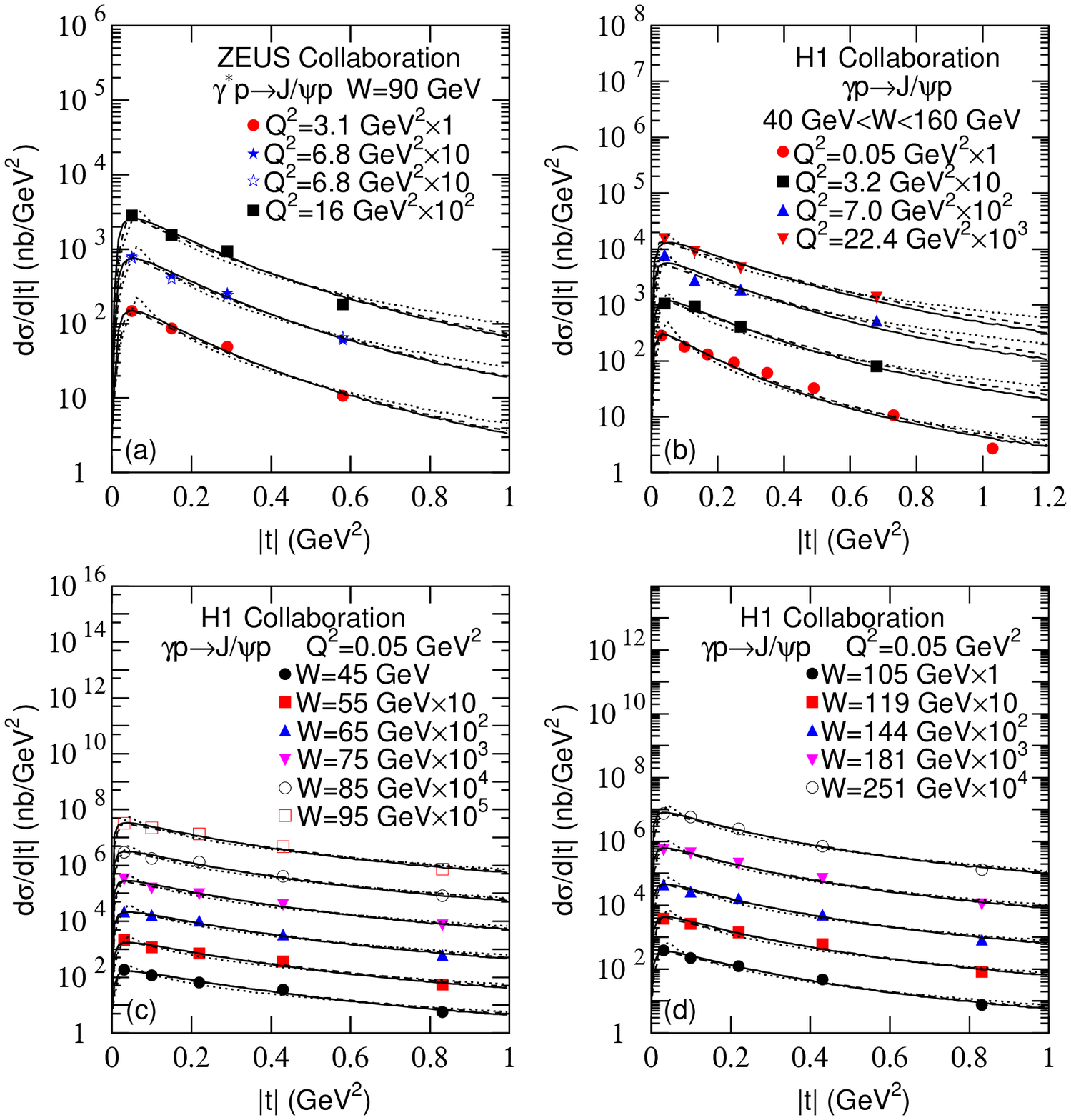}
\end{center}
\justifying\noindent {\small Fig. 3. The differential
cross-section in squared momentum transfer of (a) $\gamma^*p
\rightarrow J/\psi p$ and (b)--(d) $\gamma p \rightarrow J/\psi p$
produced in $ep$ collisions at (a) $W=90$ GeV, in (b) 40 ${\rm
GeV}<W<160$ GeV, and at (c)(d) $Q^2=0.05$ GeV$^2$. The
experimental data points from (a) exclusive and (b)--(d)
non-exclusive productions are measured by the ZEUS~\cite{21} and
H1 Collaborations~\cite{22}, respectively, with different $Q^2$
marked in panels (a) and (b), as well as with different $W$ marked
in panels (c) and (d), where in panel (a) the first and second
$Q^2=6.8$ GeV$^2$ are averaged from the ranges of $Q^2=2$--100 and
5--10 GeV$^2$, respectively. Similar to Figures 1 and 2, the data
points are fitted by the Monte Carlo calculations based on Eq.
(8).}
\end{figure*}

\begin{figure*}[!htb]
\begin{center}
\includegraphics[width=13cm]{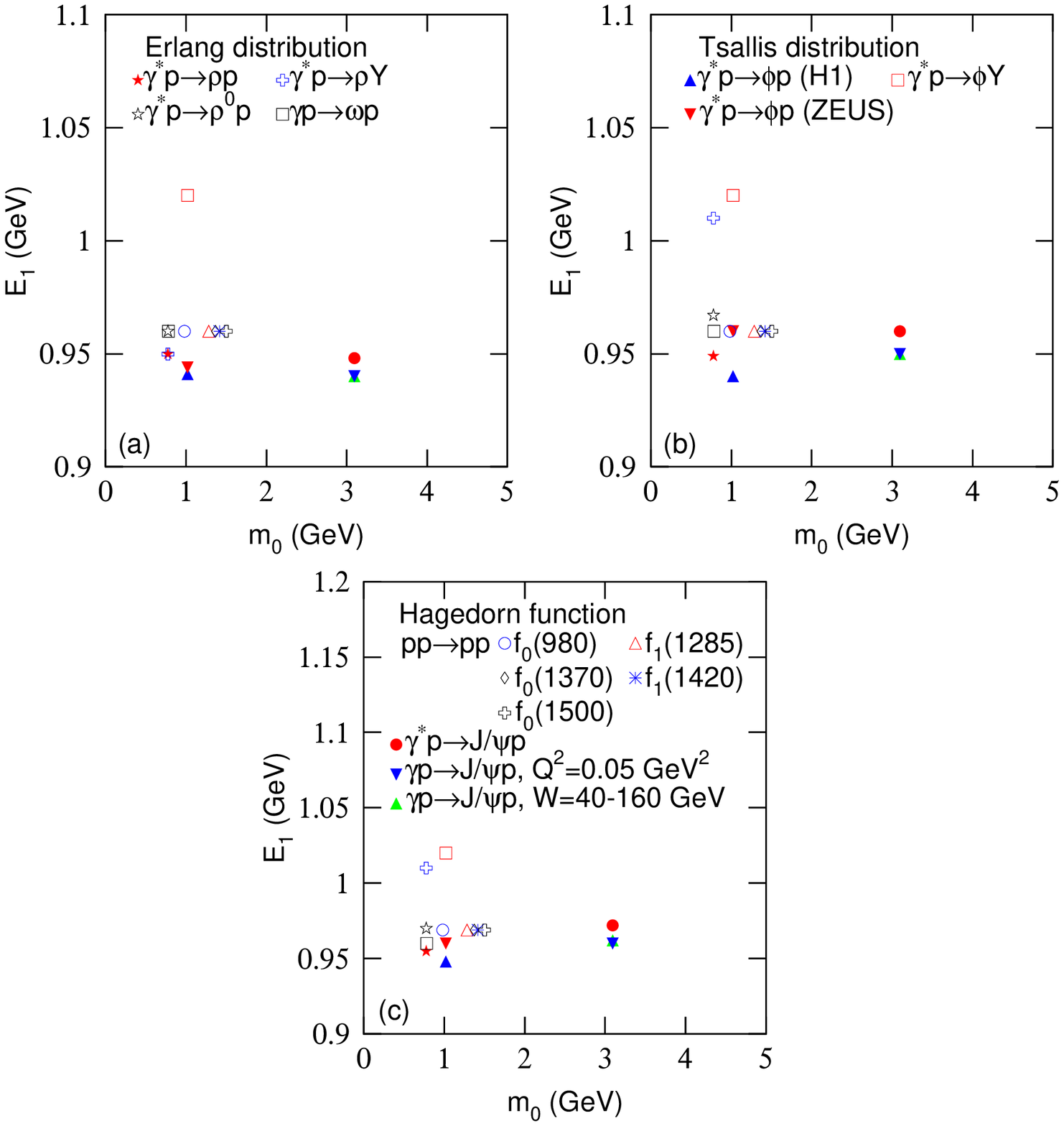}
\end{center}
\justifying\noindent {\small Fig. 4. The dependences of $E_1$ on
$m_0$ for different two-body reactions which are marked in the
panels. Panels (a)--(c) correspond to the results from the Erlang
distribution, Tsallis distribution, and Hagedorn function,
respectively.}
\end{figure*}

\begin{figure*}[htbp]
\begin{center}
\includegraphics[width=13cm]{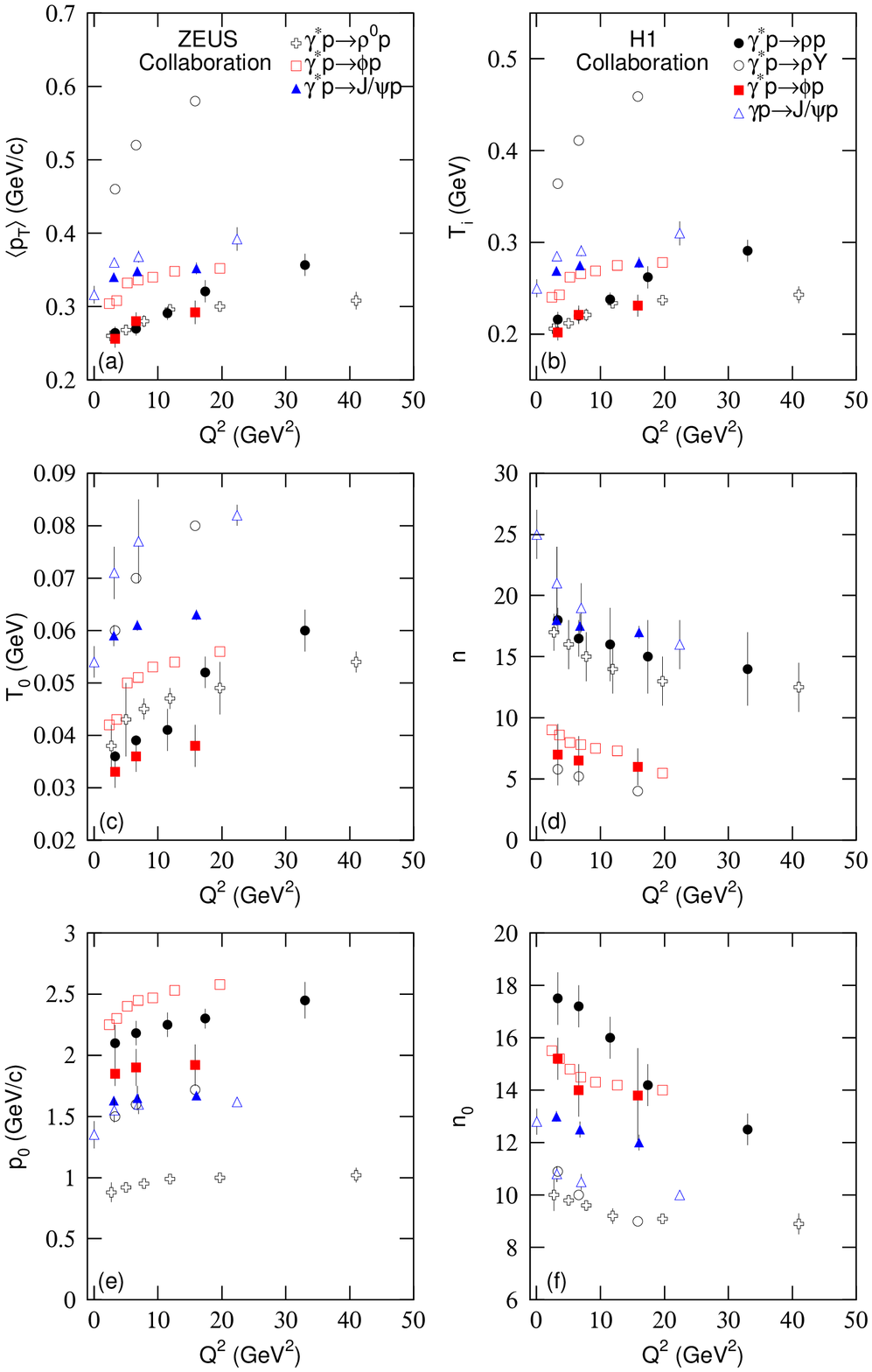}
\end{center}
\justifying\noindent {\small Fig. 5. The dependences of (a)
$\langle p_T\rangle$, (b) $T_i$, (c) $T_0$, (d) $n$, (e) $p_0$,
and (f) $n_0$ on $Q^2$ for different two-body reactions. The
symbols represent the parameter values extracted from Figures 1--3
and listed in Tables 1--3. Here, $\langle p_T\rangle=n_s\langle
p_t\rangle$ from Table 1 and $T_0=T$ from Table 2.}
\end{figure*}

\begin{figure*}[htbp]
\begin{center}
\includegraphics[width=13cm]{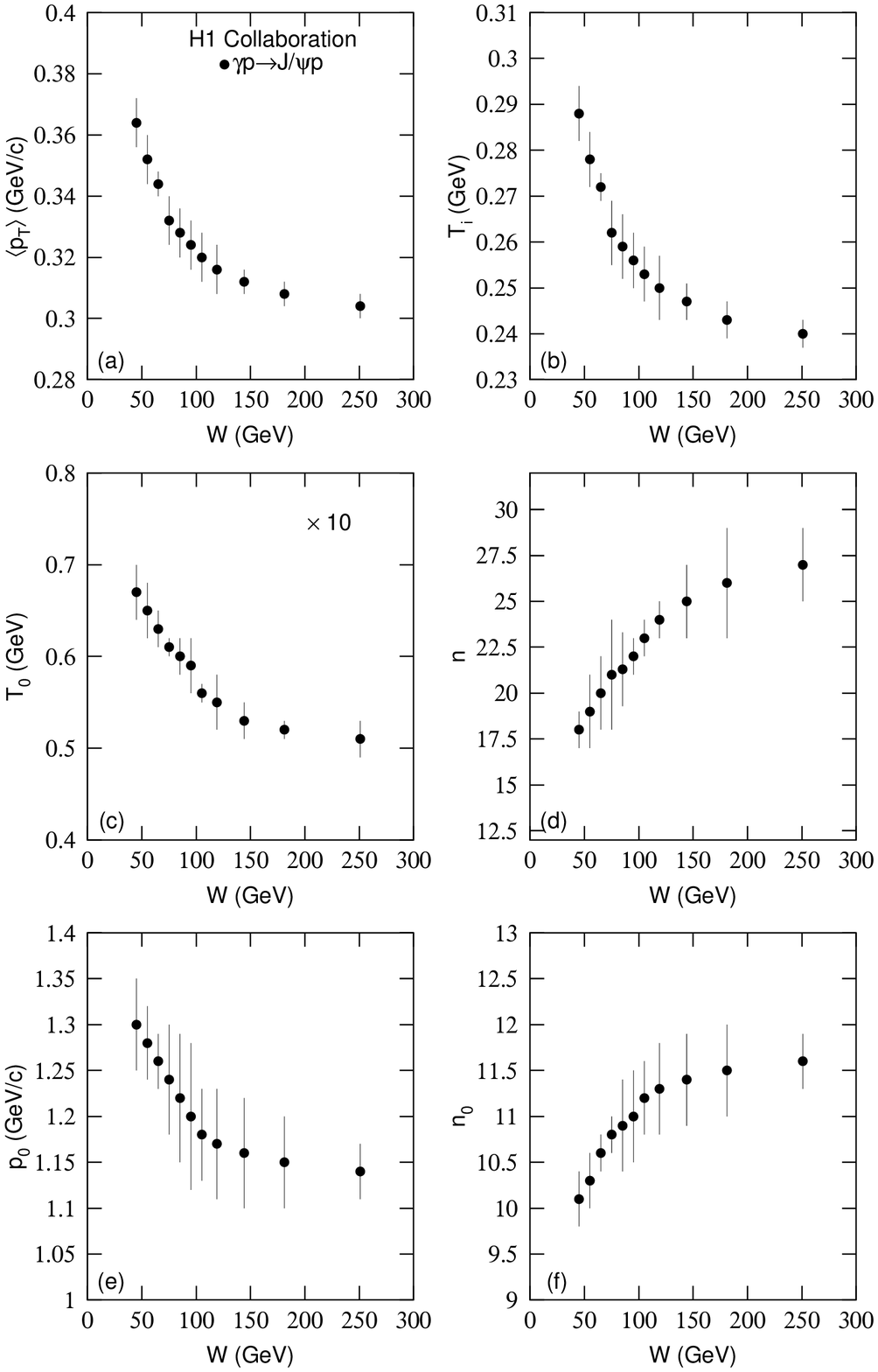}
\end{center}
\justifying\noindent {\small Fig. 6. The dependences of (a)
$\langle p_T\rangle$, (b) $T_i$, (c) $T_0$, (d) $n$, (e) $p_0$,
and (f) $n_0$ on $W$ for $\gamma p\rightarrow J/\psi p$ reactions.
The symbols represent the parameter values extracted from Figure 3
and listed in Tables 1--3. Here, $\langle p_T\rangle=n_s\langle
p_t\rangle$ from Table 1 and $T_0=T$ from Table 2.}
\end{figure*}

\begin{table*}[htbp]
\vspace{.0cm} \justifying\noindent {\small Table 1. Values of
$E_1$, $\langle p_t\rangle$, $n_s$, $T_i$, and $\chi^2$/ndof
corresponding to the solid curves in Figures 1--3, where $n_s=3$
for Figure 1(a) and $n_s=4$ for other panels, which are not listed
in the table to avoid trivialness. In some cases, ndof is less
than 1, which is denoted by $``-"$ in the last column, and the
corresponding curve is only to guide the eyes. For Figure 3(a),
the first and second $Q^2=6.8$ GeV$^2$ are averaged from the
ranges of $Q^2=2$--100 and 5--10 GeV$^2$, respectively.}
\vspace{-4mm}
\begin{center}
{\small
\begin{tabular} {cccccccc}\\ \hline\hline Figure & Reaction & Main selection & $E_1$ (GeV)  & $\langle p_t\rangle$ (GeV/$c$) & $T_i$ (GeV) & $\chi^2$/ndof\\
\hline
Figure 1(a) & $\gamma^*p \rightarrow \rho p$   & $Q^2=3.3$ GeV$^2$        & $0.950\pm0.004$ & $0.088\pm0.003$ & $0.216\pm0.008$ & $18/3$\\
            &                                  & 6.6                      & $0.950\pm0.001$ & $0.090\pm0.003$ & $0.220\pm0.008$ & $5/3$\\
            &                                  & 11.5                     & $0.950\pm0.002$ & $0.097\pm0.003$ & $0.238\pm0.007$ & $4/3$\\
            &                                  & 17.4                     & $0.950\pm0.002$ & $0.107\pm0.005$ & $0.262\pm0.012$ & $7/3$\\
            &                                  & 33.0                     & $0.950\pm0.002$ & $0.119\pm0.005$ & $0.291\pm0.012$ & $4/3$\\
Figure 1(b) & $\gamma^*p \rightarrow \rho Y$   & $Q^2=3.3$ GeV$^2$        & $0.950\pm0.002$ & $0.115\pm0.002$ & $0.364\pm0.007$ & $28/7$\\
            &                                  & 6.6                      & $0.950\pm0.010$ & $0.130\pm0.004$ & $0.411\pm0.013$ & $11/7$\\
            &                                  & 15.8                     & $0.950\pm0.010$ & $0.145\pm0.010$ & $0.459\pm0.032$ & $22/7$\\
Figure 1(c) & $\gamma^*p \rightarrow \rho^0p$  & $Q^2=2.7$ GeV$^2$        & $0.960\pm0.001$ & $0.065\pm0.001$ & $0.206\pm0.003$ & $17/1$\\
            &                                  & 5.0                      & $0.960\pm0.003$ & $0.067\pm0.001$ & $0.212\pm0.003$ & $3/1$\\
            &                                  & 7.8                      & $0.960\pm0.003$ & $0.070\pm0.002$ & $0.221\pm0.007$ & $11/1$\\
            &                                  & 11.9                     & $0.960\pm0.002$ & $0.074\pm0.001$ & $0.234\pm0.003$ & $1/1$\\
            &                                  & 19.7                     & $0.960\pm0.001$ & $0.075\pm0.001$ & $0.237\pm0.003$ & $3/1$\\
            &                                  & 41.0                     & $0.960\pm0.003$ & $0.077\pm0.003$ & $0.243\pm0.009$ & $3/1$\\
\hline
Figure 2(a) & $\gamma p \rightarrow \omega p$  & 70 ${\rm GeV}<W<90$ GeV  & $0.960\pm0.003$ & $0.046\pm0.002$ & $0.145\pm0.006$ & $4/2$\\
Figure 2(b) & $\gamma^*p \rightarrow \phi p$   & $Q^2=3.3$ GeV$^2$        & $0.941\pm0.001$ & $0.064\pm0.003$ & $0.202\pm0.009$ & $11/3$\\
            &                                  & 6.6                      & $0.941\pm0.003$ & $0.070\pm0.003$ & $0.221\pm0.010$ & $5/3$\\
            &                                  & 15.8                     & $0.941\pm0.003$ & $0.073\pm0.004$ & $0.231\pm0.012$ & $10/3$\\
            & $\gamma^*p \rightarrow \phi Y$   & $W=75$ GeV               & $1.020\pm0.001$ & $0.100\pm0.005$ & $0.316\pm0.016$ & $1/-$\\
Figure 2(c) & $\gamma^*p \rightarrow \phi p$   & $Q^2=2.4$ GeV$^2$        & $0.944\pm0.001$ & $0.076\pm0.001$ & $0.240\pm0.003$ & $1/-$\\
            &                                  & 3.6                      & $0.944\pm0.001$ & $0.077\pm0.001$ & $0.243\pm0.004$ & $1/-$\\
            &                                  & 5.2                      & $0.944\pm0.001$ & $0.083\pm0.001$ & $0.262\pm0.004$ & $1/-$\\
            &                                  & 6.9                      & $0.944\pm0.001$ & $0.084\pm0.001$ & $0.266\pm0.004$ & $1/-$\\
            &                                  & 9.2                      & $0.944\pm0.002$ & $0.085\pm0.001$ & $0.269\pm0.003$ & $1/-$\\
            &                                  & 12.6                     & $0.944\pm0.005$ & $0.087\pm0.003$ & $0.275\pm0.009$ & $9/-$\\
            &                                  & 19.7                     & $0.944\pm0.001$ & $0.088\pm0.001$ & $0.278\pm0.003$ & $2/-$\\
Figure 2(d) & $pp \rightarrow ppf_0(980)$      & $\sqrt{s_{NN}}=29.1$ GeV & $0.960\pm0.001$ & $0.079\pm0.001$ & $0.250\pm0.003$ & $5/2$\\
            & $pp \rightarrow ppf_1(1285)$     &                          & $0.960\pm0.005$ & $0.068\pm0.001$ & $0.215\pm0.003$ & $6/2$\\
            & $pp \rightarrow ppf_0(1370)$     &                          & $0.960\pm0.004$ & $0.070\pm0.004$ & $0.221\pm0.013$ & $60/2$\\
            & $pp \rightarrow ppf_1(1420)$     &                          & $0.960\pm0.003$ & $0.079\pm0.003$ & $0.250\pm0.010$ & $7/2$\\
            & $pp \rightarrow ppf_0(1500)$     &                          & $0.960\pm0.001$ & $0.080\pm0.001$ & $0.253\pm0.003$ & $13/2$\\
\hline
Figure 3(a) & $\gamma^*p \rightarrow J/\psi p$ & $Q^2=3.1$ GeV$^2$        & $0.948\pm0.004$ & $0.085\pm0.001$ & $0.269\pm0.003$ & $2/-$\\
            &                                  & 6.8                      & $0.948\pm0.002$ & $0.087\pm0.001$ & $0.275\pm0.003$ & $5/-$\\
            &                                  & 6.8                      & $0.948\pm0.002$ & $0.087\pm0.001$ & $0.275\pm0.003$ & $3/-$\\
            &                                  & 16.0                     & $0.948\pm0.001$ & $0.088\pm0.002$ & $0.278\pm0.006$ & $7/-$\\
Figure 3(b) & $\gamma p \rightarrow J/\psi p$  & $Q^2=0.05$ GeV$^2$       & $0.940\pm0.002$ & $0.079\pm0.003$ & $0.250\pm0.010$ & $80/-$\\
            &                                  & 3.2                      & $0.940\pm0.002$ & $0.090\pm0.001$ & $0.285\pm0.003$ & $2/-$\\
            &                                  & 7.0                      & $0.940\pm0.002$ & $0.092\pm0.002$ & $0.291\pm0.006$ & $10/-$\\
            &                                  & 22.4                     & $0.940\pm0.002$ & $0.098\pm0.004$ & $0.310\pm0.013$ & $3/4$\\
Figure 3(c) & $\gamma p \rightarrow J/\psi p$  & $W=45$ GeV               & $0.940\pm0.002$ & $0.091\pm0.002$ & $0.288\pm0.006$ & $11/1$\\
            &                                  & 55                       & $0.940\pm0.002$ & $0.088\pm0.002$ & $0.278\pm0.006$ & $14/1$\\
            &                                  & 65                       & $0.940\pm0.002$ & $0.086\pm0.001$ & $0.272\pm0.003$ & $6/1$\\
            &                                  & 75                       & $0.940\pm0.001$ & $0.083\pm0.002$ & $0.262\pm0.007$ & $15/1$\\
            &                                  & 85                       & $0.940\pm0.002$ & $0.082\pm0.002$ & $0.259\pm0.007$ & $13/1$\\
            &                                  & 95                       & $0.940\pm0.002$ & $0.081\pm0.002$ & $0.256\pm0.006$ & $9/1$\\
Figure 3(d) & $\gamma p \rightarrow J/\psi p$  & $W=105$ GeV              & $0.940\pm0.002$ & $0.080\pm0.002$ & $0.253\pm0.006$ & $7/1$\\
            &                                  & 119                      & $0.940\pm0.002$ & $0.079\pm0.002$ & $0.250\pm0.007$ & $21/1$\\
            &                                  & 144                      & $0.940\pm0.001$ & $0.078\pm0.001$ & $0.247\pm0.004$ & $10/1$\\
            &                                  & 181                      & $0.940\pm0.002$ & $0.077\pm0.001$ & $0.243\pm0.004$ & $14/1$\\
            &                                  & 251                      & $0.940\pm0.002$ & $0.076\pm0.001$ & $0.240\pm0.003$ & $3/1$\\
\hline
\end{tabular}}
\end{center}
\end{table*}

\begin{table*}[htbp]
\vspace{.0cm} \justifying\noindent {\small Table 2. Values of
$E_1$, $T$, $n$, and $\chi^2$/ndof corresponding to the dashed
curves in Figures 1--3, where $``-"$ in the last column denotes
the case of ${\rm ndof}<1$ and the corresponding curve is only to
guide the eyes.} \vspace{-4mm}
\begin{center}
{\small
\begin{tabular} {ccccccccc}\\ \hline\hline Figure & Reaction & Main selection & $E_1$ (GeV)  & $T$ (GeV) & $n$ & $\chi^2$/ndof\\
\hline
Figure 1(a) & $\gamma^*p \rightarrow \rho p$   & $Q^2=3.3$ GeV$^2$        & $0.949\pm0.001$ & $0.036\pm0.001$ & $18.0\pm1.0$    & $20/3$\\
            &                                  & 6.6                      & $0.949\pm0.002$ & $0.039\pm0.001$ & $16.5\pm1.5$    & $4/3$\\
            &                                  & 11.5                     & $0.949\pm0.001$ & $0.041\pm0.004$ & $16.0\pm3.0$    & $4/3$\\
            &                                  & 17.4                     & $0.949\pm0.003$ & $0.052\pm0.003$ & $15.0\pm3.0$    & $4/3$\\
            &                                  & 33.0                     & $0.949\pm0.002$ & $0.060\pm0.004$ & $14.0\pm3.0$    & $4/3$\\
Figure 1(b) & $\gamma^*p \rightarrow \rho Y$   & $Q^2=3.3$ GeV$^2$        & $1.010\pm0.003$ & $0.060\pm0.005$ & $5.8\pm0.4$     & $34/7$\\
            &                                  & 6.6                      & $1.010\pm0.003$ & $0.070\pm0.004$ & $5.2\pm0.4$     & $13/7$\\
            &                                  & 15.8                     & $1.010\pm0.002$ & $0.080\pm0.008$ & $4.0\pm1.0$     & $24/7$\\
Figure 1(c) & $\gamma^*p \rightarrow \rho^0p$  & $Q^2=2.7$ GeV$^2$        & $0.967\pm0.003$ & $0.038\pm0.004$ & $17.0\pm1.5$    & $25/1$\\
            &                                  & 5.0                      & $0.967\pm0.001$ & $0.043\pm0.007$ & $16.0\pm2.0$    & $16/1$\\
            &                                  & 7.8                      & $0.967\pm0.001$ & $0.045\pm0.002$ & $15.0\pm2.0$    & $11/1$\\
            &                                  & 11.9                     & $0.967\pm0.002$ & $0.047\pm0.002$ & $14.0\pm2.0$    & $4/1$\\
            &                                  & 19.7                     & $0.967\pm0.003$ & $0.049\pm0.005$ & $13.0\pm2.0$    & $10/1$\\
            &                                  & 41.0                     & $0.967\pm0.001$ & $0.054\pm0.002$ & $12.5\pm2.0$    & $2/1$\\
\hline
Figure 2(a) & $\gamma p \rightarrow \omega p$  & 70 ${\rm GeV}<W<90$ GeV  & $0.960\pm0.001$ & $0.021\pm0.001$ & $20.0\pm4.0$  & $3/2$\\
Figure 2(b) & $\gamma^*p \rightarrow \phi p$   & $Q^2=3.3$ GeV$^2$        & $0.940^{+0.003}_{-0.002}$ & $0.033\pm0.003$ & $7.0\pm2.5$  & $6/3$\\
            &                                  & 6.6                      & $0.940\pm0.002$ & $0.036\pm0.003$ & $6.5\pm2.0$   & $5/3$\\
            &                                  & 15.8                     & $0.940\pm0.002$ & $0.038\pm0.004$ & $6.0\pm1.5$   & $15/3$\\
            & $\gamma^*p \rightarrow \phi Y$   & $W=75$ GeV               & $1.020\pm0.003$ & $0.065\pm0.008$ & $5.8\pm1.7$   & $2/-$\\
Figure 2(c) & $\gamma^*p \rightarrow \phi p$   & $Q^2=2.4$ GeV$^2$        & $0.960\pm0.001$ & $0.042\pm0.001$ & $9.0\pm0.7$ & $1/-$\\
            &                                  & 3.6                      & $0.960\pm0.001$ & $0.043\pm0.001$ & $8.6\pm0.8$ & $1/-$\\
            &                                  & 5.2                      & $0.960\pm0.001$ & $0.050\pm0.002$ & $8.0\pm0.5$ & $1/-$\\
            &                                  & 6.9                      & $0.960\pm0.001$ & $0.051\pm0.001$ & $7.8\pm0.4$ & $1/-$\\
            &                                  & 9.2                      & $0.960\pm0.001$ & $0.053\pm0.001$ & $7.5\pm0.5$ & $2/-$\\
            &                                  & 12.6                     & $0.960\pm0.001$ & $0.054\pm0.001$ & $7.3\pm0.9$ & $7/-$\\
            &                                  & 19.7                     & $0.960\pm0.001$ & $0.056\pm0.001$ & $5.5\pm0.5$ & $1/-$\\
Figure 2(d) & $pp \rightarrow ppf_0(980)$      & $\sqrt{s_{NN}}=29.1$ GeV & $0.960\pm0.002$ & $0.051\pm0.003$ & $8.0\pm1.2$  & $20/2$\\
            & $pp \rightarrow ppf_1(1285)$     &                          & $0.960\pm0.003$ & $0.040\pm0.003$ & $12.6\pm1.0$ & $11/2$\\
            & $pp \rightarrow ppf_0(1370)$     &                          & $0.960\pm0.003$ & $0.042\pm0.003$ & $12.0\pm3.0$ & $78/2$\\
            & $pp \rightarrow ppf_1(1420)$     &                          & $0.960\pm0.003$ & $0.050\pm0.001$ & $9.5\pm0.5$  & $7/2$\\
            & $pp \rightarrow ppf_0(1500)$     &                          & $0.960\pm0.002$ & $0.051\pm0.001$ & $7.0\pm0.5$  & $18/2$\\
\hline
Figure 3(a) & $\gamma^*p \rightarrow J/\psi p$ & $Q^2=3.1$ GeV$^2$        & $0.960\pm0.002$ & $0.059\pm0.002$ & $18.0\pm0.5$ & $4/-$\\
            &                                  & 6.8                      & $0.960\pm0.002$ & $0.061\pm0.001$ & $17.5\pm1.0$ & $5/-$\\
            &                                  & 6.8                      & $0.960\pm0.002$ & $0.061\pm0.001$ & $17.5\pm1.0$ & $1/-$\\
            &                                  & 16.0                     & $0.960\pm0.002$ & $0.063\pm0.002$ & $17.0\pm0.5$ & $9/-$\\
Figure 3(b) & $\gamma p \rightarrow J/\psi p$  & $Q^2=0.05$ GeV$^2$       & $0.950\pm0.010$ & $0.054\pm0.003$ & $25.0\pm2.0$ & $88/-$\\
            &                                  & 3.2                      & $0.950\pm0.010$ & $0.071\pm0.005$ & $21.0\pm3.0$ & $4/-$\\
            &                                  & 7.0                      & $0.950\pm0.003$ & $0.077\pm0.008$ & $19.0\pm2.0$ & $5/-$\\
            &                                  & 22.4                     & $0.950\pm0.003$ & $0.082\pm0.002$ & $16.0\pm2.0$ & $1/4$\\
Figure 3(c) & $\gamma p \rightarrow J/\psi p$  & $W=45$ GeV               & $0.950\pm0.004$ & $0.067\pm0.003$ & $18.0\pm1.0$ & $9/1$\\
            &                                  & 55                       & $0.950\pm0.003$ & $0.065\pm0.003$ & $19.0\pm2.0$ & $11/1$\\
            &                                  & 65                       & $0.950\pm0.005$ & $0.063\pm0.002$ & $20.0\pm2.0$ & $9/1$\\
            &                                  & 75                       & $0.950\pm0.003$ & $0.061\pm0.001$ & $21.0\pm3.0$ & $6/1$\\
            &                                  & 85                       & $0.950\pm0.002$ & $0.060\pm0.002$ & $21.3\pm2.0$ & $9/1$\\
            &                                  & 95                       & $0.950\pm0.001$ & $0.059\pm0.003$ & $22.0\pm1.0$ & $8/1$\\
Figure 3(d) & $\gamma p \rightarrow J/\psi p$  & $W=105$ GeV              & $0.950\pm0.001$ & $0.056\pm0.001$ & $23.0\pm1.0$ & $5/1$\\
            &                                  & 119                      & $0.950\pm0.005$ & $0.055\pm0.003$ & $24.0\pm1.0$ & $16/1$\\
            &                                  & 144                      & $0.950\pm0.003$ & $0.053\pm0.002$ & $25.0\pm2.0$ & $9/1$\\
            &                                  & 181                      & $0.950\pm0.004$ & $0.052\pm0.001$ & $26.0\pm3.0$ & $19/1$\\
            &                                  & 251                      & $0.950\pm0.002$ & $0.051\pm0.002$ & $27.0\pm2.0$ & $9/1$\\
\hline
\end{tabular}}
\end{center}
\end{table*}

\begin{table*}[htbp]
\vspace{.0cm} \justifying\noindent {\small Table 3. Values of
$E_1$, $p_0$, $n_0$, and $\chi^2$/ndof corresponding to the dotted
curves in Figures 1--3, where $``-"$ in the last column denotes
the case of ${\rm ndof}<1$ and the corresponding curve is only to
guide the eyes.} \vspace{-4mm}
\begin{center}
{\small
\begin{tabular} {ccccccccc}\\ \hline\hline Figure & Reaction & Main selection & $E_1$ (GeV)  & $p_0$ (GeV/$c$) & $n_0$ & $\chi^2$/ndof\\
\hline
Figure 1(a) & $\gamma^*p \rightarrow \rho p$   & $Q^2=3.3$ GeV$^2$        & $0.955\pm0.001$ & $2.10\pm0.15$ & $17.5\pm1.0$ & $36/3$\\
            &                                  & 6.6                      & $0.955\pm0.001$ & $2.18\pm0.10$ & $17.2\pm0.8$ & $10/3$\\
            &                                  & 11.5                     & $0.955\pm0.001$ & $2.25\pm0.10$ & $16.0\pm0.8$ & $10/3$\\
            &                                  & 17.4                     & $0.955\pm0.001$ & $2.30\pm0.08$ & $14.2\pm0.8$ & $12/3$\\
            &                                  & 33.0                     & $0.955\pm0.001$ & $2.45\pm0.15$ & $12.5\pm0.6$ & $6/3$\\
Figure 1(b) & $\gamma^*p \rightarrow \rho Y$   & $Q^2=3.3$ GeV$^2$        & $1.010\pm0.003$ & $1.50\pm0.11$ & $10.9\pm0.2$ & $40/7$\\
            &                                  & 6.6                      & $1.010\pm0.004$ & $1.60\pm0.13$ & $10.0\pm0.5$ & $17/7$\\
            &                                  & 15.8                     & $1.010\pm0.004$ & $1.72\pm0.15$ & $9.0\pm0.5$  & $18/7$\\
Figure 1(c) & $\gamma^*p \rightarrow \rho^0p$  & $Q^2=2.7$ GeV$^2$        & $0.970\pm0.002$ & $0.88\pm0.08$ & $10.0\pm0.6$ & $69/1$\\
            &                                  & 5.0                      & $0.970\pm0.001$ & $0.92\pm0.04$ & $9.8\pm0.2$  & $30/1$\\
            &                                  & 7.8                      & $0.970\pm0.001$ & $0.95\pm0.02$ & $9.6\pm0.2$  & $22/1$\\
            &                                  & 11.9                     & $0.970\pm0.001$ & $0.99\pm0.04$ & $9.2\pm0.3$  & $20/1$\\
            &                                  & 19.7                     & $0.970\pm0.002$ & $1.00\pm0.03$ & $9.1\pm0.1$  & $15/1$\\
            &                                  & 41.0                     & $0.970\pm0.001$ & $1.02\pm0.06$ & $8.9\pm0.4$  & $2/1$\\
\hline
Figure 2(a) & $\gamma p \rightarrow \omega p$  & 70 ${\rm GeV}<W<90$ GeV  & $0.960\pm0.004$ & $1.28\pm0.08$ & $17.0\pm1.5$ & $6/2$\\
Figure 2(b) & $\gamma^*p \rightarrow \phi p$   & $Q^2=3.3$ GeV$^2$        & $0.948\pm0.001$ & $1.85\pm0.10$ & $15.2\pm0.8$ & $3/3$\\
            &                                  & 6.6                      & $0.948\pm0.003$ & $1.90\pm0.15$ & $14.0\pm1.0$ & $10/3$\\
            &                                  & 15.8                     & $0.948\pm0.010$ & $1.92\pm0.17$ & $13.8\pm1.8$ & $25/3$\\
            & $\gamma^*p \rightarrow \phi Y$   & $W=75$ GeV               & $1.020\pm0.002$ & $1.85\pm0.10$ & $11.5\pm0.5$ & $4/-$\\
Figure 2(c) & $\gamma^*p \rightarrow \phi p$   & $Q^2=2.4$ GeV$^2$        & $0.960\pm0.001$ & $2.25\pm0.11$ & $15.5\pm0.8$ & $5/-$\\
            &                                  & 3.6                      & $0.960\pm0.001$ & $2.30\pm0.11$ & $15.2\pm0.5$ & $3/-$\\
            &                                  & 5.2                      & $0.960\pm0.001$ & $2.40\pm0.25$ & $14.8\pm1.1$ & $4/-$\\
            &                                  & 6.9                      & $0.960\pm0.001$ & $2.45\pm0.13$ & $14.5\pm0.7$ & $4/-$\\
            &                                  & 9.2                      & $0.960\pm0.001$ & $2.47\pm0.13$ & $14.3\pm0.8$ & $4/-$\\
            &                                  & 12.6                     & $0.960\pm0.001$ & $2.53\pm0.05$ & $14.2\pm0.3$ & $4/-$\\
            &                                  & 19.7                     & $0.960\pm0.001$ & $2.58\pm0.06$ & $14.0\pm0.5$ & $1/-$\\
Figure 2(d) & $pp \rightarrow pp f_0(980)$     & $\sqrt{s_{NN}}=29.1$ GeV & $0.969\pm0.001$ & $2.22\pm0.03$ & $14.7\pm0.2$ & $24/2$\\
            & $pp \rightarrow pp f_1(1285)$    &                          & $0.969\pm0.002$ & $2.05\pm0.18$ & $18.5\pm1.5$ & $21/2$\\
            & $pp \rightarrow pp f_0(1370)$    &                          & $0.969\pm0.001$ & $2.25\pm0.36$ & $19.0\pm2.0$ & $120/2$\\
            & $pp \rightarrow pp f_1(1420)$    &                          & $0.969\pm0.001$ & $2.45\pm0.21$ & $18.0\pm1.0$ & $16/2$\\
            & $pp \rightarrow pp f_0(1500)$    &                          & $0.969\pm0.001$ & $2.70\pm0.20$ & $17.9\pm1.0$ & $30/2$\\
\hline
Figure 3(a) & $\gamma^*p \rightarrow J/\psi p$ & $Q^2=3.1$ GeV$^2$        & $0.972\pm0.001$ & $1.63\pm0.03$ & $13.0\pm0.2$ & $8/-$\\
            &                                  & 6.8                      & $0.972\pm0.002$ & $1.65\pm0.10$ & $12.5\pm0.3$ & $14/-$\\
            &                                  & 6.8                      & $0.972\pm0.002$ & $1.65\pm0.10$ & $12.5\pm0.3$ & $3/-$\\
            &                                  & 16.0                     & $0.972\pm0.003$ & $1.67\pm0.03$ & $12.0\pm0.3$ & $17/-$\\
Figure 3(b) & $\gamma p \rightarrow J/\psi p$  & $Q^2=0.05$ GeV$^2$       & $0.962\pm0.001$ & $1.35\pm0.11$ & $12.8\pm0.5$ & $195/-$\\
            &                                  & 3.2                      & $0.962\pm0.001$ & $1.55\pm0.10$ & $10.8\pm0.3$ & $8/-$\\
            &                                  & 7.0                      & $0.962\pm0.001$ & $1.60\pm0.08$ & $10.5\pm0.3$ & $4/-$\\
            &                                  & 22.4                     & $0.962\pm0.001$ & $1.62\pm0.01$ & $10.0\pm0.1$ & $1/4$\\
Figure 3(c) & $\gamma p \rightarrow J/\psi p$  & $W=45$ GeV               & $0.960\pm0.002$ & $1.30\pm0.05$ & $10.1\pm0.3$ & $20/1$\\
            &                                  & 55                       & $0.960\pm0.001$ & $1.28\pm0.04$ & $10.3\pm0.3$ & $28/1$\\
            &                                  & 65                       & $0.960\pm0.001$ & $1.26\pm0.03$ & $10.6\pm0.2$ & $23/1$\\
            &                                  & 75                       & $0.960\pm0.001$ & $1.24\pm0.06$ & $10.8\pm0.2$ & $20/1$\\
            &                                  & 85                       & $0.960\pm0.002$ & $1.22\pm0.07$ & $10.9\pm0.5$ & $25/1$\\
            &                                  & 95                       & $0.960\pm0.001$ & $1.20\pm0.08$ & $11.0\pm0.5$ & $22/1$\\
Figure 3(d) & $\gamma p \rightarrow J/\psi p$  & $W=105$ GeV              & $0.960\pm0.001$ & $1.18\pm0.05$ & $11.2\pm0.4$ & $18/1$\\
            &                                  & 119                      & $0.960\pm0.002$ & $1.17\pm0.06$ & $11.3\pm0.5$ & $36/1$\\
            &                                  & 144                      & $0.960\pm0.001$ & $1.16\pm0.06$ & $11.4\pm0.5$ & $26/1$\\
            &                                  & 181                      & $0.960\pm0.001$ & $1.15\pm0.05$ & $11.5\pm0.5$ & $45/1$\\
            &                                  & 251                      & $0.960\pm0.001$ & $1.14\pm0.03$ & $11.6\pm0.3$ & $21/1$\\
\hline
\end{tabular}}
\end{center}
\end{table*}

Figure 2 presents the differential cross-section in squared
momentum transfer, $d\sigma/d|t|$, of (a) $\gamma p \rightarrow
\omega p$, (b) $\gamma^*p \rightarrow \phi p$ and $\gamma^*p
\rightarrow \phi Y$, (c) $\gamma^*p \rightarrow \phi p$, and (d)
$pp \rightarrow ppV$ ($V=f_0(980)$, $f_1(1285)$, $f_0(1370)$,
$f_1(1420)$, and $f_0(1500)$) produced in (a)--(c) $ep$ and (d)
$pp$ collisions in (a) 70 ${\rm GeV}<W<90$ GeV, at (b)(c) $W=75$
GeV, and at (d) proton-proton center-of-mass energy per nucleon
pair $\sqrt{s_{NN}}=29.1$ GeV. The experimental data points from
(a)(c) exclusive, (b) non-exclusive, and (d) exclusive productions
are measured by the ZEUS~\cite{17,18}, H1~\cite{15}, and WA102
Collaborations~\cite{19,20}, respectively, with different $Q^2$
for only panels (b) $Q^2=3.3$, 5, 6.6, and 15.8 GeV$^2$ and (c)
$Q^2=2.4$, 3.6, 5.2, 6.9, 9.2, 12.6, and 19.7 GeV$^2$. Similar to
Figure 1, the data points are fitted by the Monte Carlo
calculations based on Eq. (8). The values of parameters are listed
in Tables 1, 2, or 3 with $\chi^2$/ndof. One can see that in most
cases the calculations based on Eq. (8) with Eqs. (2)--(4) for
$p_{3T}$ can fit approximately the experimental data measured by
the H1 and ZEUS Collaborations.

Figure 3 displays the differential cross-section in squared
momentum transfer, $d\sigma/d|t|$, of (a) $\gamma^*p \rightarrow
J/\psi p$ and (b)--(d) $\gamma p \rightarrow J/\psi p$ produced in
$ep$ collisions at (a) $W=90$ GeV, in (b) 40 ${\rm GeV}<W<160$
GeV, and at (c)(d) $Q^2=0.05$ GeV$^2$. The experimental data
points from (a) exclusive and (b)--(d) non-exclusive productions
are measured by the ZEUS~\cite{21} and H1
Collaborations~\cite{22}, respectively, with (a) $Q^2=3.1$, 6.8
averaged in 2--100, 6.8 averaged in 5--10, and 16 GeV$^2$ and (b)
$Q^2=0.05$, 3.2, 7.0, and 22.4 GeV$^2$, as well as with (c)
$W=45$, 55, 65, 75, 85, and 95 GeV and (d) $W=105$, 119, 144, 181,
and 251 GeV. Similar to Figures 1 and 2, the data points are
fitted by the Monte Carlo calculations based on Eq. (8). The
values of parameters are listed in Tables 1, 2, or 3 with
$\chi^2$/ndof. One can see that in most cases the calculations
based on Eq. (8) with Eqs. (2)--(4) for $p_{3T}$ can fit
approximately the experimental data measured by the H1 and ZEUS
Collaborations.

From the above comparisons, we see that some fits have large
$\chi^2$ compared to ndof, corresponding to low confidence levels.
The parameters obtained from these fits are not representing the
data well. We would like to say here that these values are used
only for the qualitative description of the data tendencies, but
not the quantitative interpretation of the data size. In some
cases, ${\rm ndof}<1$, which means that there were at least as
many parameters as data points. This means that a perfect fit
should have been found. However, this was not the case here. The
reason is that we have used given functions, but not any function
such as a polynomial.

\subsection{Tendency of parameters}

The dependences of energy $E_1$ of particle 1 on rest mass $m_0$
of particle 3 for different two-body reactions are given in Figure
4, where panels (a)--(c) correspond to the results from the Erlang
distribution, Tsallis distribution, and Hagedorn function,
respectively. The types of reactions are marked in the panels.
Different symbols represent the results from different reactions
or collaborations. One can see that the production of particle 3
with larger $m_0$ does not need the participation of particle 1
with larger $E_1$.

The tendency of $E_1$ versus $m_0$ presented in Figure 4 has less
fluctuation due to the given collision energy. The results from
the three distributions or functions are almost the same, if not
equal to each other, due to the same experimental data considered.
In fact, $E_1$ should be a fixed value for given reaction in the
present work. However, because different selections such as
different $Q^2$ and $W$ are used in experiments, $E_1$ has a
slight shift from the peak value. Thus, we may regard $E_1$ as a
parameter and obtain it from the fits.

The dependences of (a) $\langle p_T\rangle$, (b) $T_i$, (c) $T_0$,
(d) $n$, (e) $p_0$, and (f) $n_0$ on average squared photon
virtuality $Q^2$ for different two-body reactions are shown in
Figure 5. The types of reactions are marked in the panels.
Different symbols for different reactions represent the parameter
values extracted from Figures 1--3 and listed in Tables 1--3,
where the Erlang distribution, Tsallis distribution, and Hagedorn
function in the ranges of available data are used. In particular,
$\langle p_T\rangle=n_s\langle p_t\rangle$ from Table 1 and
$T_0=T$ from Table 2. One can see that $\langle p_T\rangle$,
$T_i$, $T_0$, and $p_0$ increase generally with an increases in
$Q^2$, and $n$ and $n_0$ decrease significantly with increasing of
$Q^2$.

Because of $Q^2$ being a reflection of hard scale of reaction,
this is natural that a harder scale results in a higher excitation
degree and then a larger $\langle p_T\rangle$, $T_i$, and $T_0$.
In most cases, one can see a large enough $n$ or $n_0$. This means
that $q$ is close to 1 and the reaction systems stay in an
approximate equilibrium state. At harder scale, the degree of
equilibrium decreases due to more disturbance to the equilibrated
residual partons in target particle. Then, one has larger $q$ and
smaller $n$ or $n_0$ when compared with those at softer scale.

Figure 6 shows the the excitation functions of related parameters,
i.e., the dependences of (a) $\langle p_T\rangle$, (b) $T_i$, (c)
$T_0$, (d) $n$, (e) $p_0$, and (f) $n_0$ on photon-proton
center-of-mass energy $W$ for $\gamma p\rightarrow J/\psi p$
reactions. The symbols represent the parameter values extracted
from Figure 3 and listed in Tables 1--3. Again, $\langle
p_T\rangle=n_s\langle p_t\rangle$ from Table 1 and $T_0=T$ from
Table 2. One can see that $\langle p_T\rangle$, $T_i$, $T_0$, and
$p_0$ decrease with an increase in $W$, and $n$ and $n_0$ increase
with increasing of $W$.

In $\gamma p\rightarrow J/\psi p$ reactions, at higher
center-of-mass energy, the incident photon has higher energy.
Although the emitted $J/\psi$ also has higher energy, it is more
inclined to have smaller angle. As a comprehensive result, the
transverse momentum of $J/\psi$ is smaller, and then $T_i$ and
$T_0$, which are obtained from the transverse momentum are also
smaller. In addition, larger $n$ and $n_0$ at higher collision
energy means more equilibrium due to shorter collision time, and
then less disturbance to the equilibrated residual partons in
target particle. This situation is different from nucleus-nucleus
collisions in which cold or spectator nuclear effect have to be
considered.

In fact, in nucleus-nucleus collisions, secondary cascade
collisions may happen among produced particles and spectator
nucleons. The secondary collisions may cause the emission angle to
increase, and then the transverse momentum to increase. The effect
of secondary collisions is more obvious or nearly saturated at
higher energy. In nucleus-nucleus collisions at lower energy, the
system approaches equilibrium more easily due to longer
interaction time. Conversely, at higher energy, the system does
not approach equilibrium more easily due to shorter interaction
time for secondary collisions.

\subsection{Further discussion}

Before summary and conclusions, we would like to point out that
the concept of temperature used in the present work is valid.
Generally, the concept of temperature is used in a large system
with multiple particles, which stays in an equilibrium state or
approximate (local) equilibrium state. From the macroscopic point
of view, the systems of $\gamma^*p$, $\gamma p$, and $pp$
reactions are indeed small. However, we know that there are lots
of events under the same condition in the experiments. These
events obey the law of grand canonical ensemble in which the
concept of temperature is applicable.

Because the same experimental condition is used in statistics,
lots of events are in equilibrium if they consist of a large
statistical system which can be described by the grand canonical
ensemble. Particles in the large statistical system obey the same
distribution law such as the same transverse momentum
distribution. From the statistical point of view, particle
productions in high energy collisions are a statistical behavior,
and the temperature reflects the width of distribution. The higher
the temperature is, the wider the distribution is.

The temperature is also a reflection of the average kinetic energy
based on large statistical system or a single particle. For single
particle, if the distribution law of kinetic energies or
transverse momenta is known, the temperature of emission source or
interacting system is known, where the source or system means the
large thermal source from the ensemble. Generally, we say the
temperature of source or system, not saying the temperature of a
given particle, from the point of view of statistical significance
of temperature. Based on the temperature, we may compare the
experimental spectra of different particles in different
experiments.

However, different methods have used different distributions or
functions, i.e. different ``thermometers". To unify these
``thermometers" or to find transformations among them, one has to
perform quite extensive analysis. Although one may use as far as
possible the standard distribution such as the Boltzmann,
Fermi-Dirac, or Bose-Einstein distribution to fit the experimental
spectra, it is regretful that a single standard distribution
cannot fit the experimental spectra very well in general.
Naturally, one may use a two-, three-, or even multi-component
standard distribution to fit the experimental spectra, though more
parameters are introduced.

In fact, the two-, three-, or multi-component standard
distribution can be fitted satisfactorily by the Tsallis
distribution with $q>1$, because the standard distribution is
narrower than the Tsallis distribution~\cite{70}. In particular,
the standard distribution is equivalent to the Tsallis
distribution with $q=1$. It is natural to use the Tsallis
distribution to replace the standard distribution. That is, one
may use the Tsallis distribution with $q>1$ to fit the
experimental spectra and obtain the temperature, though the
Tsallis temperature is less than the standard one.

As mentioned in the first section and discussed above, some
distributions applied in large collision system can be also
applied in small collision system due to the universality,
similarity or common characteristics existing in high energy
collisions~\cite{14a,14b,14c,14d,14e,14f,14g,14h,14i,14j}. Based
on the same reason, some statistical or hydrodynamic models
applied in large system should be also applied in small system. Of
course, lots of events are needed in experiments and high
statistics is needed in calculation if performing a Monte carlo
code.

\section{Summary and conclusions}

In summary, the differential cross-section in squared momentum
transfer of $\rho$, $\rho^0$ $\omega$, $\phi$, $f_{0}(980)$,
$f_{1}(1285)$, $f_{0}(1370)$, $f_{1}(1420)$, $f_{0}(1500)$, and
$J/\psi$ produced in $\gamma^* p$, $\gamma p$, and $pp$ collisions
have been analyzed by the Monte Carlo calculations in which the
Erlang distribution, Tsallis distribution, and Hagedorn function
(inverse power-law) are separately used to describe the transverse
momentum spectra of the emitted particles. In most cases, the
model results are approximately in agreement with the experimental
data measured by the H1, ZEUS, and WA102 Collaborations. In some
cases, the fits show qualitatively the data tendencies. The values
of the initial and final-state temperatures and other related
parameters are extracted from the fitting process. The squared
photon virtuality $Q^2$ and center-of-mass energy $W$ dependent
parameters are obtained.

With an increase in $Q^2$, the quantities $\langle p_T\rangle$,
$T_i$, $T_0$, and $p_0$ increase generally, and the quantities $n$
and $n_0$ decrease significantly. $Q^2$ is a reflection of hard
scale of reaction. A harder scale results in a higher excitation
degree, and then a larger $\langle p_T\rangle$, $T_i$, and $T_0$.
In most cases, the reaction system can be regarded as an
equilibrium state. At harder scale (larger $Q^2$), the degree of
equilibrium decreases due to more disturbance to the equilibrated
residual partons in target particle, though the degree of
excitation is high.

With increasing of $W$, the quantities $\langle p_T\rangle$,
$T_i$, $T_0$, and $p_0$ decrease, and the quantities $n$ and $n_0$
increase. In $\gamma p\rightarrow J/\psi p$ reactions at high
energy, the emitted $J/\psi$ is more inclined to have small angle
and hence small $p_T$, $T_i$, and $T_0$. In addition, the system
stays in a state with higher degree of equilibrium at high energy
due to less disturbance to the equilibrated residual partons in
target particle. This situation is different from nucleus-nucleus
collisions in which the influence of cold or spectator nuclear
effect is existent.
\\
\\
\\
{\bf Data Availability}

This manuscript has no associated data or the data will not be
deposited. (Authors' comment: The data used to support the
findings of this study are included within the article and are
cited at relevant places within the text as references.)
\\
\\
{\bf Ethical Approval}

The authors declare that they are in compliance with ethical
standards regarding the content of this paper.
\\
\\
{\bf Disclosure}

The funding agencies have no role in the design of the study; in
the collection, analysis, or interpretation of the data; in the
writing of the manuscript; or in the decision to publish the
results.
\\
\\
{\bf Conflicts of Interest}

The authors declare that there are no conflicts of interest
regarding the publication of this paper.
\\
\\
{\bf Acknowledgments}

The work of Q.W. and F.H.L. was supported by the National Natural
Science Foundation of China under Grant Nos. 12047571, 11575103,
and 11947418, the Scientific and Technological Innovation Programs
of Higher Education Institutions in Shanxi (STIP) under Grant No.
201802017, the Shanxi Provincial Natural Science Foundation under
Grant No. 201901D111043, and the Fund for Shanxi ``1331 Project"
Key Subjects Construction. The work of K.K.O. was supported by the
Ministry of Innovative Development of Uzbekistan within the
fundamental project on analysis of open data on heavy-ion
collisions at RHIC and LHC.
\\

\end{document}